\documentclass[12pt]{article}
\usepackage[dvips]{graphicx}
\usepackage{latexsym}
\usepackage[abs]{overpic}
\usepackage{wrapfig}
\usepackage{amssymb}
\usepackage{amsmath}
\usepackage{color}

\input epsf
\newcommand{\skyp}[1]{}

\def\Z {\bb{Z}}

\def\rem#1{}
\def\Z{{\mathbb{Z}}} % \Z=\mathbb{Z}

\def\th{{\theta}}

\def\em{{\emptyset}}

\def\mA{\mathcal{A}}

\newcommand{\Res}[1][]{\operatorname*{\mathrm{Res}_\mathnormal{#1}}}

\def\rem#1{}

\newcommand\non{\nonumber \\}
\newcommand{\bel}{\begin{eqnarray}}
\newcommand{\ee}{\end{eqnarray}}

\makeatletter
\@addtoreset{equation}{section}
\makeatother
\begin{document}

\begin{titlepage}

\bigskip
\hfill\vbox{\baselineskip12pt
\hbox{KEK-TH-1708}
\hbox{}
}
\bigskip\bigskip\bigskip\bigskip\bigskip\bigskip\bigskip\bigskip

\begin{center}
\Large{ %SCI:(non-)factorization
Factorization  of 4d  $\mathcal{N}=1$ superconformal index
}
\end{center}

\bigskip
\bigskip
\bigskip
\bigskip
\bigskip

\centerline{  Yutaka Yoshida,}
\bigskip
\centerline{\it High Energy Accelerator Research Organization (KEK)}
\centerline{\it Tsukuba, Ibaraki 305-0801, Japan}
\centerline{ yyyyosida-AT-gmail.com}
\bigskip
\bigskip

\bigskip
\bigskip
\bigskip
\bigskip
\begin{abstract}
We study the factorization of four dimensional $\mathcal{N}=1$ superconformal index for $U(N) (SU(N))$ SQCD with 
$N_F$ fundamental and anti-fundamental chiral multiplets. 
When both 
the anomaly free R-charge  assignment  and the traceless condition for total vorticities are satisfied, we find that the superconformal index  
factorizes to a pair of the elliptic uplift of the vortex partition functions. 
We also study the relation between open topological string and the the elliptic uplift of the vortex partition functions.
In the three dimensional limit, we show index for $U(N)$ theory reduces to the 
 factorized form  of  the partition function on the three dimensional squashed sphere.  
\end{abstract}

\end{titlepage}

\newpage
\baselineskip=18pt

\tableofcontents
\section{Introduction}
The localization computation of three dimensional  $\mathcal{N} \ge 3$ supersymmetric  theories are performed in \cite{Kim:2009wb, Kapustin:2009kz}.
The results are generalized to $\mathcal{N}=2$ with general charge assignments \cite{Jafferis:2010un, Hama:2010av, Imamura:2011su, Hama:2011ea}. In the  localization calculation, the path integrals reduce to the matrix model like multi-contour integrals.

It is revealed in \cite{Pasquetti:2011fj} that $U(1)$ gauge theories on the three dimensional squashed sphere $S^3_b$ possess a remarkable vortex and anti-vortex factorization property
 by performing contour integral. 
Such  a vortex and anti-vortex factorization  is generalized to  $G=U(N)$ gauge group in \cite{Hwang:2012jh, Taki:2013opa}. See also
Abelian quiver case \cite{Chen:2013pha}.
In  general,  the partition functions of three dimensional $\mathcal{N}=2$ theories on $S^1 \times S^2$ or $S^3_b$ are believed to  be 
 factorized,  at least theory has sufficient global symmetries to make the vacua gapped.  
The fundamental building block for the partition functions for $\mathcal{N}=2$ theories 
in the three dimension is called holomorphic block \cite{Beem:2012mb}.
The difference between partition function on  $S^1 \times S^2$ and that on $S^3_b$ is only sewing
procedure and holomorphic block is the universal. 

It is natural to think that the partition function of $\mathcal{N}=1$ theories in four dimensions also possess factorization properties and 
there also exists four dimensional analog of holomorphic block which becomes fundamental building block for the partition function on $S^1 \times S^3$ or $T^2 \times S^2$. 
In this article, we present the first evidence for factorization for  $\mathcal{N}=1$ superconformal index in four dimensions. 
We consider the $U(N)$ SQCD with $N_F$-flavors fundamental chiral multiplets and anti-fundamental chiral multiplets without the superpotential.
We perform the contour integrals for the superconformal index and study relation to the vortex and anti-vortex factorization. 

This article is organized as follows. In section \ref{section2}, we introduce $\mathcal{N}=1$ the superconformal index in four dimensions.
In  section \ref{section3}, we first evaluate the contour integral for  the formal superconformal index for the $U(1)$ gauge theory  and 
show that the vortex and anti-vortex factorization only occurs when  the anomaly free R-charge charge assignment is satisfied. Next we generalize 
the calculation to non-Abelian gauge group $U(N)$. In this case, we find that the factorization only occurs when both 
the anomaly free R-charge charge assignment  and traceless condition for total vorticities are satisfied.  The vortex partition function for the 
four dimensional theory becomes elliptic (theta function) uplift of  two dimensional vortex partition function.
In   section \ref{topological}, we study the open topological string which give the elliptic uplift of vortex partition function.
In  section \ref{3dlimit}, we take the three dimensional limit and study the relation between  superconformal index and the factorized partition function on $S^3_b$.     
The section \ref{summary} is devoted to summary.

\section{$\mathcal{N}=1$ superconformal index in four dimensions}
\label{section2}
The partition function on $S^1 \times S^3$ with twisted periodic boundary condition along $S^1$ which respect supersymmetries define the 
certain BPS index. The index called superconformal index in four dimensions is introduced in \cite{Romelsberger:2005eg, Kinney:2005ej}.  
%The superconformal indices introduced in  which count particular BPS operators and make it possible to  test  IR dualities or AdS/CFT correspondence
%quantitatively.
The $\mathcal{N}=1$ superconformal index in four dimensions  is defined by
\bel
\mathcal{I}=\mathrm{tr} \left( (-1)^F e^{D-\frac{3}{2} R-2J_L} s^{2J_L+2 J_R-\frac{R}{2}} t^{2J_R-2J_L-\frac{R}{2}} \prod_{I} z^{F_I}_I \right).
\ee
Here $D$, $R$,  $J_L(J_R)$ and $F_I$ are the dilation, the R-charge, the Cartan generators of left(right) $SU(2)$ isometry of $S^3$ 
 and  the flavor charges. The superconformal index counts the BPS operators which saturate the bound $D-\frac{3}{2} R-2J_L \ge 0$.
The  superconformal index can be expressed in  multi-contour representation as
\bel
\mathcal{I}=\frac{((s;s)_{\infty}(t;t)_{\infty})^{r}}{|W|} \oint_{\mathbb{T}^r} \prod_{a=1}^r \frac{d x_a}{2\pi i x_a}  Z^{1-\text{loop}}_{\text{vec}} Z^{1-\text{loop}}_{\text{chi}}. 
\ee
Here $r$ is the rank of gauge group $G$.  $Z^{1-\text{loop}}_{\text{vec}}$ is the one-loop determinant of  the vector multiplet
and $Z^{1-\text{loop}}_{\text{chi}}$ is the one-loop determinant of the chiral multiplets. They are given by
\bel
&&Z^{1-\text{loop}}_{\text{vec}}=\prod_{\alpha > 0}  \th (e^{i \alpha(y)} ;s ) \th (e^{-i \alpha(y)} ;t ), \\
&&Z^{1-\text{loop}}_{\text{chi}}=\prod_{\rho \in R} \prod_{I} \Gamma ((st)^{\frac{R}{2}} e^{-i \rho(y)} z^{F_I}_I;s,t),
\ee
with $x_a=e^{i y_a}$.
The definition of the theta function $\th (x,q)$ and the elliptic gamma function $\Gamma (x;s,t)$ are summarized in appendix.
The integration contours are taken as unit circles.

We can also introduce FI-term for  over all  $U(1)$ factor of gauge group  
\bel
\mathcal{L}_{\text{FI}} =  \zeta \mathrm{Tr} (\frac{2i}{r_3} A_4-D),  
\label{FIterm}
\ee 
which only contributes to the saddle point value in the localization calculation.
Here $\zeta$ is the FI-parameter and $r_3$ is the radius of $S^3$. $A_4$ is the gauge field along the $S^1$ circle.
For simplicity, we omit the FI-term in the later calculation, but it is easy to recover its contribution. We consider the FI-term contribution in section \ref{3dlimit}; we study the 
relation between the superconformal index in four dimensions and  the partition function in three dimensions.  

When the gauge group is $G=U(N)$ and the  matter chiral multiplets are the $N_F$-flavors with  fundamental representation  
and $N_F$-flavors with  anti-fundamental representation, the superconformal index is 
written as
\bel
\mathcal{I}^{U(N)}_{N_F} &&= \frac{(s;s)^N (t;t)^N}{N!} 
\oint_{\mathbb{T}^N} \prod_{a=1}^{N}  \frac{d x_a}{2\pi i}  \prod_{ a  > b } 
\th (x_a x^{-1}_b ;s ) \th (x^{-1}_a x_b ;t ) \non
%\prod_{1 \le a  \neq b \le N} \Gamma (e^{i x_a -ix_b}  ;s,t ) 
&& \qquad \qquad  \prod_{a=1}^{N} \prod_{I=1}^{N_F} \Gamma ((st)^{\frac{R}{2}} x^{-1}_a z_I;s,t) \Gamma ((st)^{\frac{\tilde{R}}{2}} x_a {\tilde{z}_I};s,t). 
\label{UNindex1}
\ee 
In the next section, we evaluate the above multi-contour integrals and study the relation to the vortex  partition function 
and anti-vortex partition function  factorization.
%%%%%%%%%%%%%%%%%%%%%%%%%%%%%%%%%%%
\section{Factorization of superconformal index}
\label{section3}
\subsection{Abelian case}
In this subsection, we consider the Abelian gauge group $G=U(1)$. Since the Abelian gauge theories in the four dimensions is infrared free,
the index (\ref{UNindex1}) of this theory is the formal object, but it is useful to see the factorization property and  to generalize to non-Abelian case.
We assume that fugacities are analytically continued to the region $|pq z_I| <1$ and the residues are evaluated at the poles of the one-loop determinant of 
$N_F$ fundamental chiral multiplets. As in the case of 3d $\mathcal{N}=2$ superconformal indices \cite{Hwang:2012jh}, By shifting $z_I \to z_{I} (st)^{c_I} $, we can set $R=0$ and $\tilde{R}=0$. In the same reason, we have omitted the Baryonic charges from the beginning .
Then we evaluate residues at pole $x =z_{I'} s^{j} t^{k}, (j,k \in \Z_{\ge 0}, I=1, \cdots,N_F)$ in the  one-loop determinant of the chiral multiplets.
   
The index is written as
\bel
\mathcal{I}^{U(1)}_{N_F}=-  \sum_{I'=1}^{N_F}    \sum_{j,k=0}^{\infty}   (s;s)_{\infty}(t;t)_{\infty}  \Res_{x=z_{I'} t^{j} t^{k}}   \prod_{I=1}^{N_F} \Gamma ( x^{-1} z_I;s,t) \Gamma ( x {\tilde{z}_I};s,t). 
\ee
The contributions from one-loop determinant of the  fundamental chiral multiplet are given by
\bel
\prod_{I=1  \atop  I \neq I'}^{N_F} \Gamma ( x^{-1} z_I ;s,t) \Big|_{x=z_{I'} s^{j} t^{k}}
&&=\prod_{I=1 \atop I  \neq I'}^{N_F} \Gamma (  s^{-j} t^{-k}  z^{-1}_{I'} z_I ;s,t) \non
&& = \prod_{I=1  \atop  I \neq I'}^{N_F}  (-z^{-1}_{I'} z_I)^{-jk} s^{k\frac{j(j+1)}{2}} t^{j\frac{k(k+1)}{2}}    \non
&&\quad  \prod_{l=1}^{j} \th^{-1} (s^{-l}  z^{-1}_{I'} z_I;t) \prod_{m=1}^{k}\th^{-1} ( t^{-m} z^{-1}_{I'} z_I;s)
 \Gamma( z^{-1}_{I'} z_I;s,t). \non
 \label{Abel1}
\ee
Here we have used an identity (\ref{ellipticID}) for the elliptic gamma function.
The residue is evaluated as 
\bel
&&\Res_{y=s^{-j} t^{-k}}  \Gamma(  y; s,t) \non
&& \quad =(-y)^{jk} s^{k \frac{j(j-1)}{2}} t^{j \frac{k(k-1)}{2}} \prod_{l=0}^{j-1} \theta^{-1}(s^l y;t) \prod_{m=0}^{k-1} \theta^{-1}(t^m y;s) \Big|_{y=s^{-j} t^{-k}}
\Res_{y=s^{-j} t^{-k}}  \Gamma(s^j t^k y; s,t) \non
&& \quad = (-1)^{jk+1} s^{k \frac{j(j+1)}{2} } t^{j \frac{k(k+1)}{2}} (s;s)^{-1}_{\infty} (t;t)^{-1}_{\infty}
 \prod_{l=1}^{j} \theta^{-1}(s^{-l} ;t) \prod_{m=1}^{k} \theta^{-1}(t^{-m};s).
\label{Abel2}
\ee
From the second line to third line in (\ref{Abel2}), we have used the relation (\ref{diffth1}) and 
\bel
\Res_{y=1} \Gamma(  y; s,t) =- (s;s)^{-1}_{\infty} (t;t)^{-1}_{\infty}.
\ee 
In a similar manner, we can evaluate the contribution from the anti-fundamental chiral multiplets as
\bel
\prod_{I=1  }^{N_F} \Gamma ( y \tilde{z}_I ;s,t) \Big|_{y=z_{I'} s^{j} t^{k}}
&&=\prod_{I=1 }^{N_F} \Gamma (  s^{j} t^{k}  z_{I'} \tilde{z}_I ;s,t) \non
&&=\prod_{I=1 }^{N_F}  (-z_{I'} \tilde{z}_I)^{-jk} s^{-k \frac{j(j-1)}{2}} t^{-j \frac{k(k-1)}{2}}  \non
&&   \prod_{l=0}^{j-1} \theta(s^l z_{I'} \tilde{z}_I;t) 
\prod_{m=0}^{k-1} \theta(t^m z_{I'} \tilde{z}_I ;s) \Gamma (  z_{I'} \tilde{z}_I ;s,t). 
\label{Abelanti}
\ee
From (\ref{Abel1}), (\ref{Abel2}) and (\ref{Abelanti}), we obtain the index as 
\bel
\mathcal{I}^{U(1)}_{N_F} 
&&=\sum_{I'=1}^{N_F}  
\Bigl(  \prod_{I=1}^{N_F}  \Gamma( z_{I'} \tilde{z}_{I} ; s,t) \Bigl) \Bigl( \prod_{I=1 \atop I \neq I'}^{N_F} \Gamma( z^{-1}_{I'} z_{I};s,t) \Bigr)  
 \sum_{j,k=0}^{\infty} \left(  \prod_{I=1}^{N_F} (z_{I} \tilde{z}_I)^{-1}  st \right)^{jk} \non
&&
\Bigl(    
 \prod_{l=1}^{j} \frac{\prod_{I=1}^{N_F}  \theta (s ^{l-1} z_{I'} \tilde{z}_{I};t ) }
{    \theta (s^{-l} ;t )  \prod_{I=1 \atop I \neq I'}^{N_F}  \th (s^{-l}  z^{-1}_{I'} z_{I};t) }  \Bigr) 
 \Bigl(  
\prod_{m=1}^{k} \frac{\prod_{I=1}^{N_F}  \theta (t^{m-1} z_{I'} \tilde{z}_{I};s)}
{   \theta (t^{-m};s) \prod_{I=1 \atop I \neq I'}^{N_F}   \th ( t^{-m} z^{-1}_{I'} z_{I};s) } \Bigr).  \non
\ee
Since we have shifted fugacity as $z_{I \text{old}} \to z_{I \text{new}} (st)^{c_I}$ to set the R-charges zero, it satisfies $
 (  {z_{I} \tilde{z}_I})^{-1}_{\text{new}} st = (  z_{I} \tilde{z}_I)^{-1}_{\text{old}} (st)^{- (\frac{R}{2} + \frac{\tilde{R}}{2}) +1}$. In order to occur complete factorization, the R-charges have to  satisfy $R=\tilde{R}=1$.
In this case, we find the complete factorization occurs, because the original flavor fugacities for $SU(N_F) \times SU(N_F)$ satisfy $\prod_{I=1}^{N_F}z_{I, \text{old}}= \prod_{I=1}^{N_F} \tilde{z}_{I, \text{old}}=1$.  $R=\tilde{R}=1$ is precisely the R-charge assignments determined from the  anomaly free condition. 
In this R-charge assignments, the superconformal index has completely factorized form:
\bel
\mathcal{I}^{U(1)}_{N_F} 
&=&\sum_{I'=1}^{N_F}  
\Bigl(  \prod_{I=1}^{N_F}  \Gamma( z_{I'} \tilde{z}_{I} ; s,t) \Bigl) \Bigl( \prod_{I=1 \atop I \neq I'}^{N_F} \Gamma( z^{-1}_{I'} z_{I};s,t) \Bigr)   
 \non
&& \quad  \Biggl[ \sum_{j=0}^{\infty}    
 \prod_{l=1}^{j} \frac{\prod_{I=1}^{N_F}  \theta (s ^{l-1} z_{I'} \tilde{z}_{I};t ) }
{    \theta (s^{-l} ;t )  \prod_{I=1 \atop I \neq I'}^{N_F}  \th (s^{-l}  z^{-1}_{I'} z_{I};t) }  \Biggr] 
 \Biggl[  
\sum_{k=0}^{\infty}  \prod_{m=1}^{k} \frac{\prod_{I=1}^{N_F}  \theta (t^{m-1} z_{I'} \tilde{z}_{I};s)}
{   \theta (t^{-m};s) \prod_{I=1 \atop I \neq I'}^{N_F}   \th ( t^{-m} z^{-1}_{I'} z_{I};s) } \Biggr].  \non
\ee
Here we emphasize that $U(1)$ formal superconformal index never factorize, if the number of fundamental chiral multiplets is different from
that of anti-fundamental chiral multiplets. This is quite different structure from  two or three dimensional theories. In  two or three dimensions, the factorization
occurs when the number of fundamental chiral multiplets is different from that of anti-fundamental chiral multiplets.   

Next, we study the relation between the superconformal index and vortex partition functions \cite{Shadchin:2006yz, Dimofte:2010tz, Yoshida:2011au}. 
In the two dimensions, it is shown in \cite{Benini:2012ui, Doroud:2012xw} that  partition functions on $S^2$ factorized to vortex and anti-vortex partition functions.
In the three dimensional case,  the factorized partition functions become trigonometric (hyperbolic) uplift of 
the  vortex partition function \cite{Pasquetti:2011fj, Hwang:2012jh, Taki:2013opa}. Thus we expect the above factorized form is related to elliptic uplift of the vortex partition function.
To see this, we introduce 
 $s=e^{ i \varepsilon}, z_I=e^{i m_I}, \tilde{z}_I=e^{- i \tilde{m}_I}  $.
Then a factorized part is written as
\bel
\prod_{l=1}^{j} \frac{\prod_{I=1}^{N_F}  \theta (s^{l-1} z_{I'} \tilde{z}_{I};t ) }
{    \theta (s^{-l} ;t )  \prod_{I=1 \atop I \neq I'}^{N_F}  \th (s^{-l}  z^{-1}_{I'} z_{I};t) }  
= \prod_{l=1}^{j} \frac{\prod_{I=1}^{N_F}  \th ( e^{i \left (l-1) \varepsilon +m_{I'} -\tilde{m}_{I} \right)} ;t ) }
{    \th (e^{- i l  \varepsilon} ;t )  \prod_{I=1 \atop I \neq I'}^{N_F}  \th ( e^{i \left( -l \varepsilon  + m_{I} -m_{I'} \right)};t) }. 
\label{abelvortex}
\ee
By changing $\th (e^{i x};t) \to x$, we find that (\ref{abelvortex}) agrees with 
 the two dimensional 
$U(1)$ vortex partition function of  $N_F$ fundamental flavors and anti-fundamental flavors with vorticity $j$ at a vacuum labeled by twisted mass $m_{I'}$. 
Thus (\ref{abelvortex}) is precisely the elliptic uplift  of vortex partition function in two dimensions. In section \ref{3dlimit},  we will also show that
the elliptic uplift of the vortex partition functions reduce to the three dimensional vortex partition function by the dimensional reduction.

%%%%%%%%%%%%%%%%%%
\subsection{Non-Abelian case}

In this subsection, we generalize the result in the previous subsection to non-Abelian gauge group $U(N)$.
As in the case of three dimensions \cite{Taki:2013opa}, by use of the Cauchy determinant formula of the theta function, we find that the  poles in the vector multiplet one-loop determinant do not 
contribute to the evaluation. 
Thus, It is enough to evaluate the residues at pole $x_a=z_{I_a} s^{j_a} t^{k_a}, (a=1, \cdots, N),  j_a ,k_a \ge \mathbb{Z}_{\ge 0}$ in the chiral multiplet one-loop determinant.   
The evaluation of the residues for chiral multiplet is quite  parallel to  the Abelian case in the previous section.
The contribution of vector multiplet is as follows.
Substituting $x_a=z_{I_a} s^{j_a} t^{k_a}$ into the vector multiplet one-loop determinant,
 we obtain 
\bel
&&\prod_{a >b }\th (z_{I_a} z^{-1}_{I_b}  s^{j_a-j_b} t^{k_a-k_b};s) \th (z^{-1}_{I_a} z_{I_b}  s^{j_b-j_a} t^{k_b-k_a};t) \non
&&=\prod_{a >b } (-z_{I_a} z^{-1}_{I_b})^{j_b-j_a-k_a+k_b} s^{\frac{1}{2} (j_a-j_b)(j_a-j_b-1)} t^{\frac{1}{2} (k_b-k_a)(k_b-k_a-1)} 
 (st)^{(j_b-j_a)(k_a-k_b)} \non
&&\left( \prod_{b=1}^N  (-1)^{j_b} s^{-\frac{1}{2}j_b(j_b+1)} \right) 
\left(\prod_{a>b} s^{-j_a (j_b+1)} \right)  \left(\prod_{a,b=1}^{N}  s^{j_b (j_a+1)} \right) \non
&&\left( \prod_{a=1}^N  (-1)^{k_a} t^{-\frac{1}{2}k_a(k_a+1)} \right) 
\left(\prod_{a>b} t^{-k_b (k_a+1)} \right)  \left(\prod_{a,b=1}^{N}  t^{k_a (k_b+1)} \right) \non
&& \left(\prod_{a,b=1}^{N} \prod_{m=0}^{k_a-1}  \frac{\th (z_{I_a} z^{-1}_{I_b} t^{-m-1};s)}{\th (z_{I_a} z^{-1}_{I_b} t^{-m+k_b};s)} \right)
\left(\prod_{a > b} \th (z_{I_a} z^{-1}_{I_b};s) \right) \non
&& \left(\prod_{a,b=1}^{N} \prod_{m=0}^{j_b-1}  \frac{\th (z_{I_b} z^{-1}_{I_a} s^{-l-1};t)}{\th (z_{I_b} z^{-1}_{I_a} s^{-m+j_a};t)} \right)
\left(\prod_{a > b} \th (z_{I_b} z^{-1}_{I_a};t) \right).
\ee
Therefore superconformal index is written as
\bel
&&\mathcal{I}^{U(N)}_{N_F} \non
 &&=
\sum_{1\le I_1 < \cdots I_N \le N_F}   
\left(\prod_{a > b}  \th (z_{I_a} z^{-1}_{I_b};s) \th (z_{I_b} z^{-1}_{I_a};t) \right) 
\left( \prod_{a=1}^N \prod_{I=1}^{N_F}  \Gamma( z_{I_a} \tilde{z}_{I} ; s,t)  \right) \left( \prod_{a=1}^N \prod_{I=1 \atop I \neq I_a}^{N_F} \Gamma( z^{-1}_{I_a} z_{I};s,t) \right) \non
&& \sum_{ \{j \} ,\{ k\} =0}^{\infty} 
{\left(  \prod_{I=1}^{N_F} z_{I} \tilde{z}_I st \right)^{j_a k_a} (st)^{\sum_{a >b } (j_b-j_a)(k_a-k_b)}}  \non
&&\left( \prod_{a >b } (-z_{I_a} z^{-1}_{I_b})^{j_b-j_a} \right) 
 (-1)^{ \sum_{b=1}^N j_b}  s^{\sum_{a >b } \left( \frac{1}{2} (j_a-j_b)(j_a-j_b-1)-j_a (j_b+1) \right) -\sum_{a=1}^N\frac{1}{2}j_a(j_a+1) +\sum_{a,b=1}^N j_b (j_a+1)}
\non 
&&\Bigl(  
  \frac{ \prod_{I=1}^{N_F} \prod_{a=1}^N \prod_{l=1}^{j_a}  \theta (s ^{l-1} z_{I_a} \tilde{z}_{I};t ) }
{  \prod_{a,b=1}^{N} \prod_{l=0}^{j_b-1}  \th (z_{I_b} z^{-1}_{I_a} s^{-l+j_a};t)
  \prod_{a=1}^{N} \prod_{l=0}^{j_a-1}   \prod_{I \not\in  \mA }  \th (s^{-l}  z^{-1}_{I_a} z_{I};t) }  \Bigr) \non
&& \left( \prod_{a >b } (-z_{I_a} z^{-1}_{I_b})^{k_b-k_a} \right) 
 (-1)^{\sum_{a=1}^N  k_a} t^{\sum_{a >b } \left( \frac{1}{2} (k_b-k_a)(k_b-k_a-1)-k_b (k_a+1) \right) -\sum_{b=1}^N \frac{1}{2}k_a(k_a+1)+\sum_{a,b=1}^{N} k_a (k_b+1)  } 
  \non
&&  
\Bigl( \frac{\prod_{I=1}^{N_F} \prod_{a=1}^N \prod_{m=1}^{k_a}  \theta (t^{m-1} z_{I_a} \tilde{z}_{I};s )}
{  \prod_{a,b=1}^{N} \prod_{m=0}^{k_b-1}  \th (z_{I_b} z^{-1}_{I_a} t^{-m+k_a};s)
  \prod_{a=1}^{N} \prod_{m=0}^{k_a-1}   \prod_{I \not\in \mA }  \th (t^{-m}  z^{-1}_{I_a} z_{I};s)  } \Bigr).  
\label{UNindex}
\ee
Here we defined $\mathcal{A}:=\{ I_1, \cdots,I_N \}$. 
The elliptic uplift of the $U(N)$ vortex partition function with $a$-th $U(1)$ vorticity $j_a$ and a vacuum labeled by $\mA$ is given 
by
\bel
Z^{V} _{\{ j_a \}, \{ I_a \}}=\frac{ \prod_{I=1}^{N_F} \prod_{a=1}^N \prod_{l=1}^{j_a}  \theta (s ^{l-1} z_{I_a} \tilde{z}_{I};t ) }
{  \prod_{a,b=1}^{N} \prod_{l=0}^{j_b-1}  \th (z_{I_b} z^{-1}_{I_a} s^{-l+j_a};t)
  \prod_{a=1}^{N} \prod_{l=0}^{j_a-1}   \prod_{I \not\in  \mA}  \th (s^{-l}  z^{-1}_{I_a} z_{I};t) }, \non
  \label{nonAbelian} 
\ee
and  that of the anti-vortex partition function with $a$-th $U(1)$ vorticity $k_a$ and a vacuum labeled by $\mA$ is given by
\bel
Z^{\bar{V}} _{\{ k_a \}, \{ I_a \}}=\frac{\prod_{I=1}^{N_F} \prod_{a=1}^N \prod_{m=1}^{k_a}  \theta (t^{m-1} z_{I_a} \tilde{z}_{I};s )}
{  \prod_{a,b=1}^{N} \prod_{m=0}^{k_b-1}  \th (z_{I_b} z^{-1}_{I_a} t^{-m+k_a};s)
  \prod_{a=1}^{N} \prod_{m=0}^{k_a-1}   \prod_{I \not\in  \mA }  \th (t^{-m}  z^{-1}_{I_a} z_{I};s)  }. \non
\ee
The (\ref{UNindex})  almost factorize, but contains the non-factorized factor with respect to  the vorticity $j_a$ and the  anti-vorticity $k_a$ 
which is given by  
\bel
\left(  \prod_{I=1}^{N_F} (z_{I} \tilde{z}_I)^{-1}  st \right)^{\sum_{a=1}^N j_a k_a} (st)^{\sum_{a >b } (j_b-j_a)(k_a-k_b)}.
\ee
The non-factorizable factor is rewritten as 
\bel
&&\left(  \prod_{I=1}^{N_F} (z_{I} \tilde{z}_I)^{-1}_{\text{new}}  st \right)^{\sum_{a=1}^N  j_a k_a} (st)^{\sum_{a >b } (j_b-j_a)(k_a-k_b)} \non
&& \qquad =   (st)^{[-N_F (\frac{R}{2}+\frac{\tilde{R}}{2}) +(N_f-N) ] ( \sum_{a=1}^N j_a k_a) +(\sum_{a=1 }^N j_a) (\sum_{a=1 }^N k_a)}. 
%&&(st)^{\sum_{a >b } (j_b-j_a)(k_a-k_b)}=(st)^{-N(\sum_{a=1}^N j_a k_a) + (\sum_{a=1 }^N j_a) (\sum_{a=1 }^N k_a)}
\ee
Therefore, the conditions for the complete factorization becomes
\bel
\label{faccond1}
&&R=\tilde{R}=1-\frac{N}{N_F}, \\ 
&&\sum_{a=1 }^N j_a=0 \quad  \text{and} \quad  \sum_{a=1 }^N k_a=0.
\label{faccond2}
\ee
The condition (\ref{faccond1}) is again the  R-charges assignments determined uniquely by the  anomaly free condition for $SU(N)$ SQCD 
 with $N_F$ fundamental and anti-fundamental chiral multiplets without superpotential.
If  the traceless condition   is imposed for the gauge group $U(N)$ (namely gauge group becomes $SU(N)$) , 
the condition (\ref{faccond2}) is satisfied. When  the delta function constraint $\delta(\sum_{a=1}^{N} x_a)$ is inserted in the integrations (\ref{UNindex1}), 
these conditions are satisfied.    
 Up to the Weyl permutations, we assume that $j_a, k_a (a=1,\cdots,N-1)$ run non-negative integers and $j_N=-\sum_{a=1}^{N-1} j_a, 
k_N=-\sum_{a=1}^{N-1} k_a$ is imposed.
It is interesting  to study   the the relation between index computation and   the fact that the overall $U(1)$ factor is decoupled  in the infrared limit.

Under the condition (\ref{faccond1}) and (\ref{faccond2}), the superconformal index completely factorize as
\bel
\mathcal{I}^{SU(N)}_{N_F}=\sum_{1\le I_1 < \cdots I_N \le N_F}  Z^{1-\text{loop}}_{\text{v.H}} Z^{1-\text{loop}}_{\text{chi.H}} 
Z^{1-\text{loop}}_{\text{a.chi.H}} {Z}_{V} Z_{\bar{V}}, 
\ee 
with
\bel
\label{onevecH}
&& Z^{1-\text{loop}}_{\text{v.H}}=\prod_{a > b}  \th (z_{I_a} z^{-1}_{I_b};s) \th (z_{I_b} z^{-1}_{I_a};t), \\
\label{onechiH}
&& Z^{1-\text{loop}}_{\text{v.H}}=\prod_{a=1}^N \prod_{I=1 \atop I \neq I_a}^{N_F} \Gamma( z^{-1}_{I_a} z_{I};s,t), \\
\label{oneachiH}
&&  Z^{1-\text{loop}}_{\text{a.chi.H}}=\prod_{a=1}^N \prod_{I=1}^{N_F}  \Gamma( z_{I_a} \tilde{z}_{I} ; s,t), 
\ee
\bel
\label{BPS}
&& {Z}_{V}=\sum_{ \{j \}' }  \left( \prod_{a =1}^N z_{I_a}^{N j_a} \right) 
   s^{\sum_{a >b } \left( \frac{1}{2} (j_a-j_b)(j_a-j_b-1)-j_a (j_b+1) \right) -\sum_{a=1}^N\frac{1}{2}j^2_a }
\non 
&& \qquad \qquad  \Bigl(  
  \frac{ \prod_{I=1}^{N_F} \prod_{a=1}^N \prod_{l=1}^{j_a}  \theta (s ^{l-1} z_{I_a} \tilde{z}_{I};t ) }
{  \prod_{a,b=1}^{N} \prod_{l=0}^{j_b-1}  \th (z_{I_b} z^{-1}_{I_a} s^{-l+j_a};t)
  \prod_{a=1}^{N} \prod_{l=0}^{j_a-1}   \prod_{I \not\in  \mA }  \th (s^{-l}  z^{-1}_{I_a} z_{I};t) }  \Bigr), \\
\label{aBPS}
&&Z_{\bar{V}}  =  \sum_{ \{k \}' } \left( \prod_{a =1}^N z_{I_a}^{N k_a} \right) 
 t^{\sum_{a >b } \left( \frac{1}{2} (k_b-k_a)(k_b-k_a-1)-k_b (k_a+1) \right) -\sum_{b=1}^N \frac{1}{2}k^2_a  } 
  \non
&&  
\qquad \qquad  \Bigl( \frac{\prod_{I=1}^{N_F} \prod_{a=1}^N \prod_{m=1}^{k_a}  \theta (t^{m-1} z_{I_a} \tilde{z}_{I};s )}
{  \prod_{a,b=1}^{N} \prod_{m=0}^{k_b-1}  \th (z_{I_b} z^{-1}_{I_a} t^{-m+k_a};s)
  \prod_{a=1}^{N} \prod_{m=0}^{k_a-1}   \prod_{I \not\in  \mA }  \th (t^{-m}  z^{-1}_{I_a} z_{I};s)  } \Bigr). 
\ee 
Here we defined $\sum_{\{ j \} ' }:=\sum_{a=1}^{N-1} \sum_{j_a=0}^{\infty} $.  

We refer to the relation with {\it Higgs branch localization} first introduced in \cite{Benini:2012ui}. In the ordinary localization, the saddle point values admit
constant field (holonomy or the scalar in vector multiplet) configurations which lead to multi-contour integrals. On the other hand, in the Higgs branch localization, saddle point admit the discrete vacua of root of Higgs branch and the point like BPS (anti-BPS) equations at the north (south) pole of $S^2$.
Again, the partition function can be evaluated in the WKB approximation around the discrete vacua. In the four dimensions, after the torus compactification, the combinations of the flavor holonomies along torus on $T^2 \times \mathbb{R}^2$ play the role of twisted masses (real masses).  
Then,  (\ref{onevecH}), (\ref{onechiH}) and (\ref{oneachiH}) can be interpreted as the one-loop determinant of the vector multiplet, the fundamental chiral multiplets and anti-fundamental chiral multiplets in the Higgs branch localization, respectively.   
In the Higgs branch localization in three dimensions \cite{Fujitsuka:2013fga, Benini:2013yva},   
the point like vortices exist at the north pole of  base space $S^2$ of $S^3$ and the point like anti-vortices exist at the south pole.
Then vortex(anti-vortex) world volume becomes one-dimensional circle $S^1$ which is the circle fiber at the north (south) pole, respectively.
In the case of $S^1 \times S^3$, there is an additional trivial $S^1$-fiber and vortex  world volume becomes two dimensional torus $T^2$.
From the point like vortices, base $S^2$ can be regarded as flat space $\mathbb{R}^2$, namely it is equivalently to consider the vortex partition function
 of the $\mathcal{N}=1$ theory on $T^2 \times \mathbb{R}^2$. 
Then Kaluza-Klein momenta along the $T^2$ will provide the elliptic deformation of vortex partition function.
This is quite analogous to the Nekrasov's instanton partition function is elliptically deformed, when $\mathcal{N}=2$ supersymmetric gauge 
theories is uplifted to torus compactified  the six dimensional $\mathcal{N}=(1,0)$ theories on $T^2 \times \mathbb{R}^4$ \cite{Hollowood:2003cv}.

The matter contents we considered in this section is the electric theory in the Seiberg duality. 
When we consider the  the magnetic theory, the factorization 
again occurs, if and only if both the traceless condition and the correct R-charges assignments are satisfied.

%%%%%%%%%%%%%%%%%%%%%%%%%%%%%%%%%%%
%\newpage
\section{Relation to open topological string}
\label{topological}
\begin{figure}
\begin{center}
\includegraphics[height=5cm]{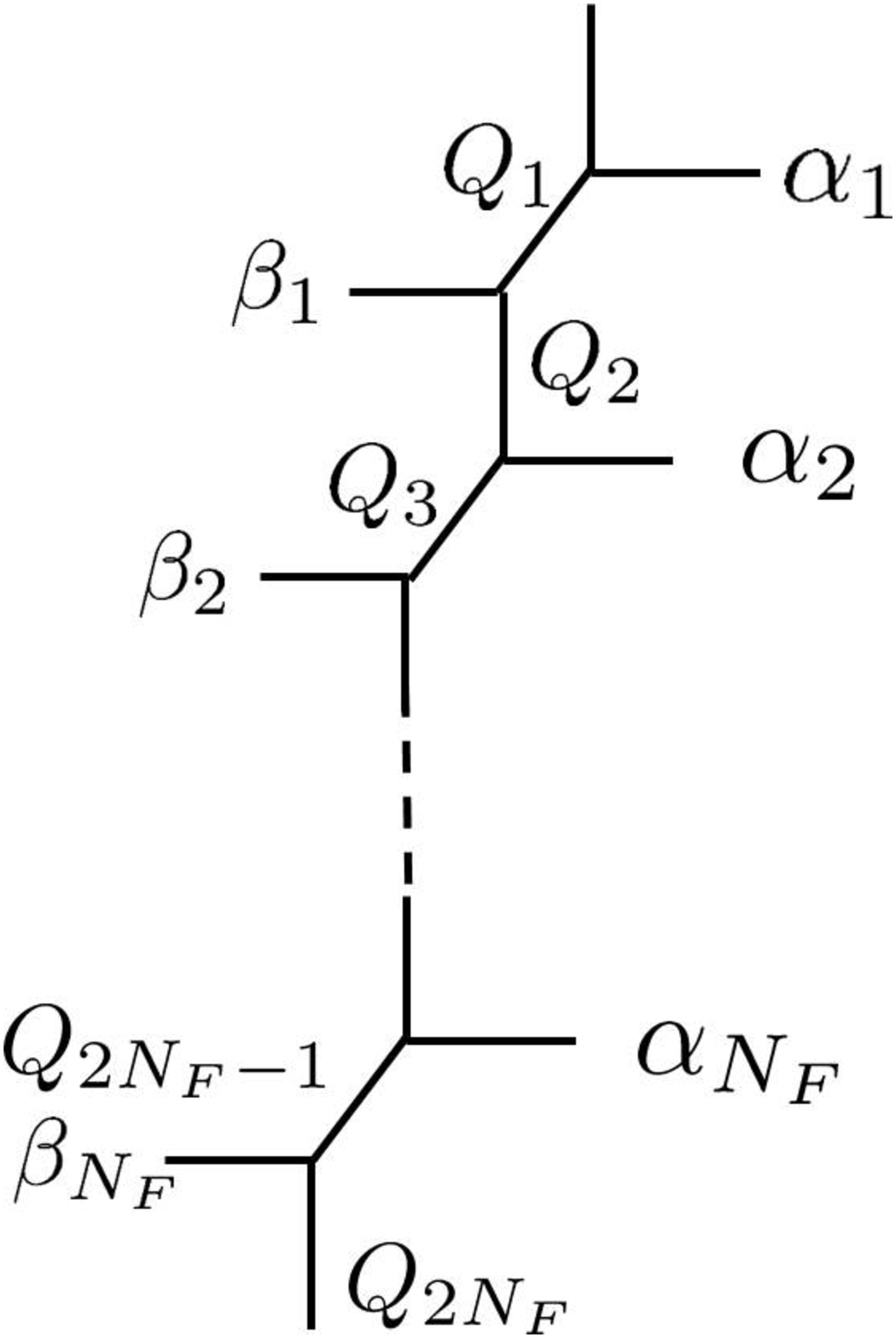}\qquad\qquad
\includegraphics[height=6cm]{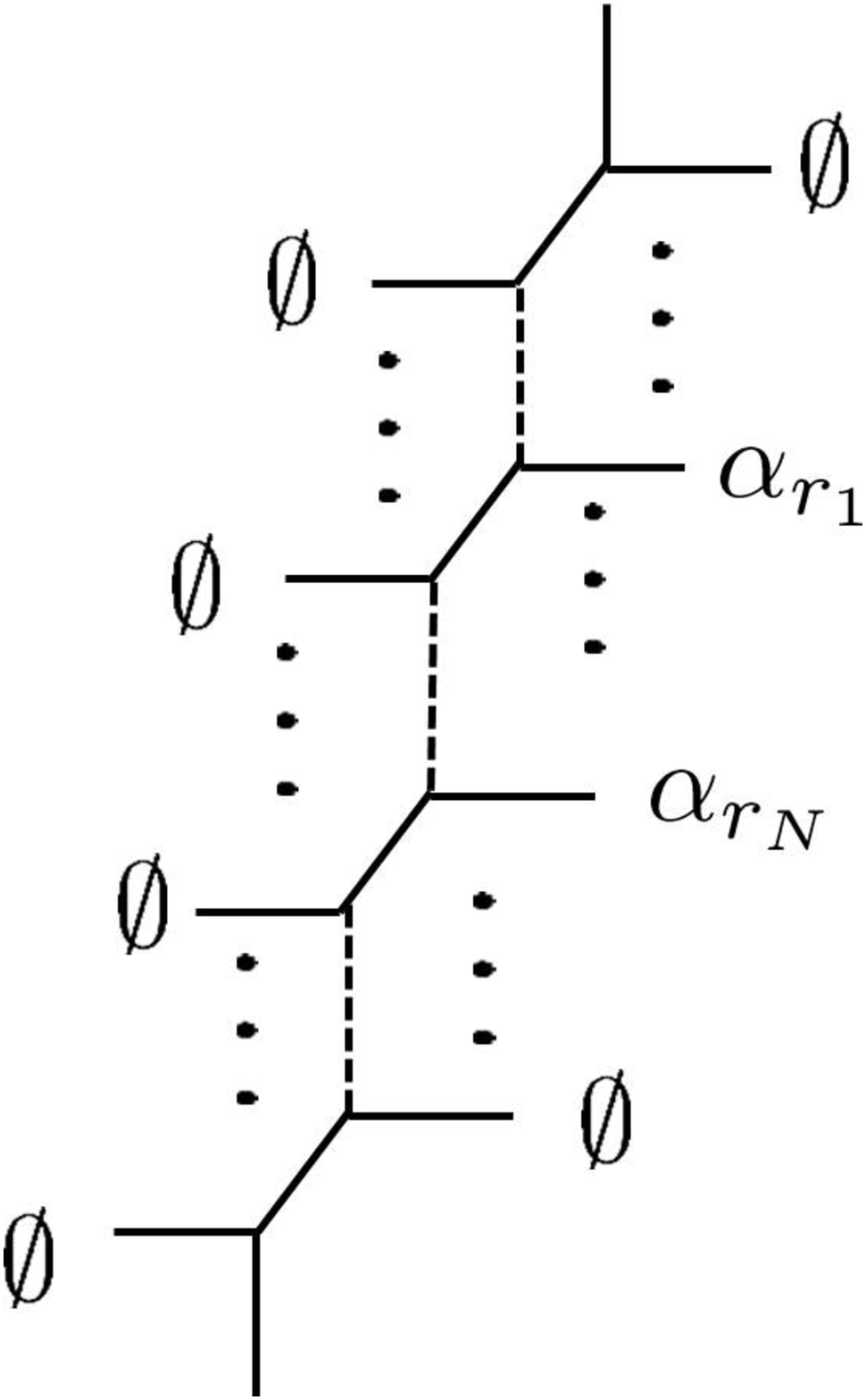}
\caption{Left: The $(p,q)$-web for the Calabi-Yau 3-folds.  Two vertical external legs are identified and becomes an internal line. $Q_a, (a=1,\cdots, 2N_F)$ represents the exponentiated Kahler parameter of 
$a$-th internal line.  Right:  a  A-brane is inserted at $r_a$-th  right horizontal external leg $(a=1, \cdots, N)$ and the representation is $1^{k_a}$ .
The other external legs are the trivial representation}
\label{fig:web}
\end{center}
\end{figure}
It is shown in \cite{Pasquetti:2011fj, Taki:2013opa} that  the  vortex partition function in the factorized partition functions coincides the open topological string partition function of strip geometry.
In this section, we study open topological string amplitude which give the elliptic uplift of  vortex partition function (\ref{nonAbelian}).
We consider the Calabi-Yau 3-fold described by  $(p,q)$-web in the figure \ref{fig:web}.
The topological vertex \cite{Aganagic:2003db} gives topological string amplitudes on this geometry. 
The (refined) topological string on the geometry we concern are studied 
  in detail \cite{Hohenegger:2013ala} \cite{Haghighat:2013tka} ( see also \cite{Hollowood:2003cv} ). 
The  open topological string amplitude for  the left of  figure \ref{fig:web} is given by
\bel
\frac{W^{\alpha_1 \cdots \alpha_{N_F}}_{\emptyset \cdots \emptyset}}{W^{\emptyset \cdots \emptyset}_{\emptyset \cdots \emptyset}}
=\left( \prod_{I=1}^{N_F} q^{\frac{||\alpha_I||^2}{2}} \tilde{Z}_{\alpha_I} (q) \right) 
\prod_{r,l=1}^{N_F} \frac{\mathcal{J}_{\em \alpha_{r+l}} (Q_{2r,2r+2l-2} ;q) 
\mathcal{J}_{\alpha_{r} \em } (Q_{2r-1, 2r+2l-3} ;q) }{\mathcal{J}_{\alpha_r \alpha_{r+l}} ( Q_{2r-1,2r+2l-2};q)}. \non
\ee
Here we defined 
\bel
\mathcal{J}_{\mu \nu}(x,q):=
\prod_{k=1}^{\infty} \prod_{(i,j) \in \mu} (1-Q^{k-1}_{U} x q^{\mu_i+\nu^{t}_i-i-j+1}) \prod_{(i,j) \in \nu} (1-Q^{k-1}_{U} x q^{-\mu^{t}_j-\nu_i+i+j-1}),
\ee
with $Q_{U}=\prod_{r=1}^{2N} Q_r$ and $Q_{a, b} =\prod_{r=a}^{b} Q_{r}$. 
The $\tilde{Z}_{\mu} (q)$ is defined by
\bel
\tilde{Z}_{\mu} (q)&=&\prod_{s \in \mu} (1-q^{l_{\mu}(s)+a_{\mu}(s)+1})^{-1},  
\ee
with $l_{\mu}(s)=\mu_i-j, a_{\mu}(s)=\mu^t_{j}-i$ for $s=(i,j)$ in the partition $\mu$.

We set the  representation of $r_a$-th right external horizontal leg  as  $\alpha_{r_{a}}=1^{k_{a}}$ for $a=1, \cdots N$ and the others as trivial representation $\alpha_{r}=\beta_r=\em$.   
Then the non-trivial $\mathcal{J}_{\mu \nu}(x,q)$ are following three types:
\bel
\mathcal{J}_{1^{k_a} \emptyset}(x,q)&=&
\prod_{k=1}^{\infty} \prod_{i=1}^{k_a} (1-x Q^{k-1}_{U} q^{1-i}), \non
\mathcal{J}_{\emptyset 1^{k_a}}(x,q)&=&
\prod_{k=1}^{\infty}  \prod_{i=1}^{k_a} (1-x Q^{k-1}_{U} q^{-1+i}), \non
\mathcal{J}_{1^{k_a} 1^{k_b}}(x,q)&=&
\prod_{k=1}^{\infty}    \prod_{i=1}^{k_a} (1-x Q^{k-1}_{U} q^{1+k_b-i})  \prod_{j=1}^{k_b} (1-x Q^{k-1}_{U} q^{-1-k_a+j}). 
\label{threetype}
\ee
It follows from (\ref{threetype}) that 
\bel
&&\prod_{r,l=1}^{N_F}  \mathcal{J}_{\em \alpha_{r+l}} (Q_{2r,2r+2l-2} ;q) 
\mathcal{J}_{\alpha_{r} \em } (Q_{2r-1, 2r+2l-3} ;q)
=\prod_{r=1}^{N_F}\prod_{a=1}^N \prod_{i=0}^{k_a-1} \th ( q^i Q_{2r_a-1,2r-1} ; Q_{U}), \non
&&\prod_{r,l=1}^{N_F} \mathcal{J}_{\alpha_r \alpha_{r+l}} ( Q_{2r-1,2r+2l-2};q) \non
&&=\left( \prod_{a,b =1}^N \prod_{i=1}^{k_a} \th ({Q}_{2r_a-1, 2r_b-2} q^{1+k_b-i} ;Q_{U}) \right)
\left( \prod_{ r \not\in \{ r_1, \cdots r_N \} } \prod_{a=1}^{N} \prod_{i=1}^{k_a} \th ({Q}_{2r_a-1, 2 r-2} q^{1-i} ;Q_{U})  \right). \non
\ee
Here we used the relation $Q^{-1}_U Q_{a, b}=Q_{b+1,a-1}$. %\bel
Therefore open topological string amplitude becomes
\bel
&&\frac{W^{\em \cdots \em 1^{k_1} \cdots 1^{k_N} \em \cdots  \em}_{\emptyset \cdots\cdots\cdots \emptyset}}{W^{\emptyset \cdots \emptyset}_{\emptyset \cdots \emptyset}} \non
&&=\left( \prod_{a=1}^{N} q^{\frac{k_a}{2}} \prod_{i=1}^{k_a} (1-q^{i})^{-1} \right) \non
&&\frac{\displaystyle \prod_{r=1}^{N_F}\prod_{a=1}^N \prod_{i=0}^{k_a-1} \th ( q^i Q_{2r_a-1,2r-1} ; Q_{U})}
{\displaystyle \left( \prod_{a,b =1}^N \prod_{i=1}^{k_a} \th ({Q}_{2r_a-1, 2r_b-2} q^{1+k_b-i} ;Q_{U}) \right)
\left( \prod_{ r \not\in \{ r_1, \cdots r_N \} } \prod_{a=1}^{N} \prod_{i=1}^{k_a} \th ({Q}_{2r_a-1, 2 r-2} q^{1-i} ;Q_{U})  \right)} \non
\ee
If we identify the parameters as
$q=s$, $Q_{U}=t$, $Q_{2I_a-1,2I-1}=z_{I_a} \tilde{z}_{I}$, $k_a=j_a$ and ${Q}_{2I_a-1, 2 I-2}=z^{-1}_{I_a} {z}_{I}$,
it  finds that the open topological string amplitude  agrees with (\ref{nonAbelian}) up to the over all  factor $\prod_{a=1}^{N} q^{\frac{k_a}{2}} \prod_{i=1}^{k_a} (1-q^{i})^{-1}$.

%%%%%%%%%%%%%%%%%%%%%%%%%%%%%%%%%
\section{Three dimensional limit}
\label{3dlimit}
When the radius of $S^1$ goes to zero, it shown in \cite{Dolan:2011rp}, \cite{Gadde:2011ia}, \cite{Imamura:2011uw} (See also \cite{Aharony:2013dha}) that 
the  superconformal indices in four dimension reduces to the partition functions on the three dimensional (squashed) sphere.  
In this section, we  consider  the three dimensional limit of $U(N)$ index (\ref{UNindex1}) and study the relation to the partition function on three dimensional squashed sphere 
\cite{Imamura:2011uw}, \cite{Imamura:2011wg}.
We set parameter as
\bel
&&s=e^{\beta \omega_1}, \quad t=e^{\beta \omega_2}, \quad \omega_1 \omega_2=1, \quad b:=\sqrt{\frac{\omega_1}{\omega_2}}, \quad Q:=b+\frac{1}{b}. \non
&&z_{I}=e^{i \beta m_I  }, \quad \tilde{z}_{I}=e^{-i \beta \tilde{m}_I }. 
\ee
Here $\beta=\frac{r_1}{r_3}$ is the ratio of the $S^1$ radius $r_1$ and   $S^3$ radius $r_3$. 
In the limit $r_1 \to 0$, the theta function and elliptic gamma function reduce to a trigonometric and the double-sin function, respectively: 
\bel
&&\lim_{\beta \to 0} \th (e^{\beta \sigma}; e^{\beta \omega_1} )  =   2  \sin  \pi b^{-1}  \sigma , \quad 
\lim_{\beta \to 0} \th (e^{\beta \sigma}; e^{\beta \omega_2} )  =   2  \sin  \pi b  {\sigma} , \\
&&\lim_{\beta \to 0} \Gamma( e^{\beta \sigma }; e^{\beta \omega_1}, e^{\beta \omega_2})
=s_b \left( i \sigma+i \frac{Q}{2} \right) 
=s^{-1}_b \left( -i \sigma -i \frac{Q}{2} \right).  
\ee
Then it follows that
\bel
\label{1loopvec}
&&\lim_{\beta \to 0} \th (z_{I_a} z^{-1}_{I_b};s) \th (z_{I_b} z^{-1}_{I_a};t)  
  = 4  \sinh \pi b^{-1} \left( m_{I_b}- m_{I_a}\right) \sinh \pi b\left( m_{I_a} - m_{I_b} \right),
\\
\label{1loopfun}
&&\lim_{\beta \to 0} \Gamma( z^{-1}_{I_a} z_{I};s,t)=  s_b \left( m_{I_a} -m_{I} +i \frac{Q}{2} \right),
\\
\label{1loopanti}
&&\lim_{\beta \to 0} \Gamma( z_{I_a} \tilde{z}_{I} ; s,t)=s^{-1}_b \left(  m_{I} -\tilde{m}_{I}  -i \frac{Q}{2} \right),
\ee
and 
\bel
&&\lim_{\beta \to 0} Z^{V} _{\{ j_a \}, \{ I_a \}} \non
&&=
\frac{ \prod_{I=1}^{N_F} \prod_{a=1}^N \prod_{l=1}^{j_a} 2  \sinh \pi b \left( m_{I_a} -\tilde{m}_{I}  -i (l-1)b \right)}
  {\displaystyle \prod_{a,b=1}^{N} \prod_{l=0}^{j_b-1} 2 \sinh \pi b  \left(  m_{I_b} -m_{I_a} +i (l-j_a) b \right) 
\prod_{a=1}^{N} \prod_{l=0}^{j_a-1}   \prod_{I \not\in  \mathcal{A} } 2 \sinh \pi b \left(  m_{I}-m_{I_a}  + il b \right) }, \non
\label{3dvortex}
\ee
\bel
&&\lim_{\beta \to 0} Z^{\bar{V}} _{\{ k_a \}, \{ I_a \}} \non
&&=
\frac{ \prod_{I=1}^{N_F} \prod_{a=1}^N \prod_{l=1}^{k_a} 2  \sinh \pi b^{-1} \left( m_{I_a} -\tilde{m}_{I}  -i (l-1)b^{-1} \right)}
  {\displaystyle \prod_{a,b=1}^{N} \prod_{l=0}^{k_b-1} 2 \sinh \pi b^{-1}  \left(  m_{I_b} -m_{I_a} +i (l-k_a) b^{-1} \right) 
\prod_{a=1}^{N} \prod_{l=0}^{k_a-1}   \prod_{I \not\in  \mathcal{A} } 2 \sinh \pi b^{-1} \left(  m_{I}-m_{I_a}  + il b^{-1} \right) }. \non
\label{3dantivortex}
\ee
We find that the right hand sides of (\ref{1loopvec}), (\ref{1loopfun})  and   (\ref{1loopanti})
correctly reproduce the one-loop determinant of the  three dimensional vector multiplet, 
the $N_F$-flavors fundamental chiral multiplets and $N_F$-flavors anti-fundamental chiral multiplets in the Higgs branch localization, respectively.
Moreover, (\ref{3dvortex})  and (\ref{3dantivortex})  also agree the vortex and anti-vortex partition functions on the squashed sphere, respectively.  
Next we consider the contribution of FI-parameter (\ref{FIterm}).
We substitute  the saddle point value at the $x_a=z_{I_a} s^{j_a} t^{k_a}$ to the $e^{i A_a}$ 
\bel
\lim_{\beta \to 0} \exp ( -S_{F I})=e^{-4 \pi^2 r^2_3 \zeta \sum_a (  m_{I_a} +b j_a+ b^{-1}  k_a)}
\ee
This also correctly reproduces the FI-term contribution in the three dimensions.
Therefore, in  the three dimensional limit,  we find that the index of  $U(N)$ theory in four dimensions directly 
reduces to the factorized partition function of $U(N)$ theory with $N_F$-flavors fundamental chiral multiples and $N_F$-flavors anti-fundamental chiral multiples
on the three dimensional squashed sphere.

%%%%%%%%%%%%%%%%%%%%%%%%%%%%%%%%%%%
\section{Summary}
\label{summary}
In this article, we have studied  factorization properties of $\mathcal{N}=1$ superconformal index in the four dimensions. 
In the $U(1)$ case, the index factorize, when the following two conditions are satisfied: the number of fundamental and anti-fundamental multiplets is same, 
the correct R-charge assignments are 
satisfied. In the factorized form, the elliptic uplift of the vortex partition function and anti-vortex function appear.
In the non-Abelian case, it found that the superconformal index completely factorize to the elliptic uplift of the vortex partition function and anti-vortex function,  only when   the traceless condition in addition to the above two condition  is  satisfied. 
We comment on several future directions.
\begin{itemize}
\item
{\it Higgs branch localization. }

In the direct contour integral evaluation,  It is obscure the reason why the vortex and anti-vortex  partition functions appear.   
Higgs branch localization directly explains them. It is interesting to perform the Higgs branch localization in four dimensions and explain 
the origin  of the parameters in the factorized form.

\item
{\it The relation between the elliptic uplift of the vortex partition and $\mathcal{N}=(0,2)$ elliptic genus.}

The instanton partition functions on $T^2 \times \mathbb{R}^4$ is equivariant elliptic genera of instanton moduli space. 
In the case of $\mathcal{N}=1$ supersymmetric theories in four dimensions,  it is known that  vortex world volume preserves $\mathcal{N}=(0,2)$ 
supersymmetry in two dimensions \cite{Edalati:2007vk}. Thus, the elliptic uplift of the vortex partition should be 
$\mathcal{N}=(0,2)$ elliptic genera of vortex moduli space. The $\mathcal{N}=(0,2)$ flavored elliptic genus is introduced in \cite{Benini:2013nda}, \cite{Benini:2013xpa} and localization is studied in detail. It is interesting to derived the elliptic uplift of vortex partition function directly from
 elliptic genus of vortex moduli space via localization.

\item
{\it Factorization of partition functions on $T^2 \times S^2$ or $S^1 \times S^3/\mathbb{Z}_n$. }

Recently, a partition function on $T^2 \times S^2$ is studied in \cite{Closset:2013vra, Closset:2013sxa}. From the Higgs branch localization perspective,  
partition function on $T^2 \times S^2$ should  also factorize. Because, point like vortices (anti-vortices ) exist on $S^2$ and the trivial $T^2$-fiber exists.

In \cite{Imamura:2013qxa}, the factorization of partition functions on $S^3/\mathbb{Z}_{n}$ is studied. 
This suggest that  the partition function on $S^1 \times S^3/\mathbb{Z}_n$ \cite{Benini:2011nc} can be regarded $S^1$ uplift of $S^3/\mathbb{Z}_{n}$  also factorized into two parts. Because we have found  that $U(N)$ theory correctly reduces to the factorized partition function in three dimensions.

In the case of $S^1 \times S^3/\mathbb{Z}_n$,  $S^1$-fiber is $\mathbb{Z}_{n}$ orbifolded. 
Then the vortex world volume becomes $S^1 \times S^1/ \mathbb{Z}_n$. This means that $\mathcal{N}=(0,2)$ elliptic genus of vortex moduli space is orbifolded
 elliptic genus and twisted sectors appear.  The twisted sectors are combined to  reproduce a 
weak  Jacobi form or modular covariance in the Landau-Ginzburg orbifolds \cite{Kawai:1993jk, Kawai:1994np}. It is interesting to 
study the modular properties of the orbifolded elliptic genus of vortex moduli space.

\item
{\it  Holomorphic blocks in four dimensions}

To construct general theory of holomorphic blocks in four dimensions is one of the most interesting and challenging future directions. 
In three dimensions, the partition function on $S^1 \times S^2$ is constructed from 
identity fusion of holomorphic blocks and  the partition function on $S^3_b$ is constructed from $S$-fusion. The holomorphic block
is universal in these two spaces. 
We conjecture that the partition functions on $T^2 \times S^2$ is identity fusion of holomorphic block in four dimensions and
the partition functions on $S^1 \times S^3$ is $S$-fusion of holomorphic block in four dimensions: 
\bel
&&Z_{T^2 \times S^2} =\sum_{\mathcal{A} }  || \mathcal{B}^{\mathcal{A}} (s,t) ||_{\text{id}}=
\sum_{\mathcal{A}} \mathcal{B}^{\mathcal{A}} (\tilde{s}, \tilde{t})
 \mathcal{B}^{\mathcal{A}} (\tilde{s}, \tilde{t}) \\ 
&&Z_{S^1 \times S^3} = \sum_{\mathcal{A}} || \mathcal{B}^{\mathcal{A}} (s,t) ||_S = \sum_{\mathcal{A}} \mathcal{B}^{\mathcal{A}} (s,t)
 \mathcal{B}^{\mathcal{A}} (t,s) 
\ee 
Here $\sum_{\mathcal{A}}$ runs all the possible choice of vacua up to the Weyl permutations.  $\tilde{s}, \tilde{t}$ in the identity fusion mean that
 $\tilde{s}=e^{2\pi i \tau }, \tilde{t}=e^{2\pi i \sigma}$. $(\tau, \sigma)$ parametrizes complex structure of $T^2 \times S^2$.

\end{itemize}

\subsection*{Acknowledgment}
%%%%%%%%%%%%%%%%%%%%%%%
%%%%%%%%%%%%%%%%%%%%%%%
%%%%%%%%%%%%%%%%%%%%%%%
The author is grateful to Masashi Fujitsuka, Masazumi Honda, Ryuichiro Kitano, Takahiro Nishinaka and Seiji Terashima for useful discussions 
and comments. 
%$ \backslash$ ( $\hat{}$ o $\hat{}$ ) /. 
On this occasion, the author  would express his thanks to several collaborators for patiently waiting for his late correspondence,  
when the  author could not take time to study due to part-time jobs to earn his daily rice. 
Finally, I am appreciate my parents for their constant support,  when I was suffered from a chronic disease. 

%%%%%%%%%%%%%%%%%%%%%%%%%%%%
\appendix
\section{Conventions and useful formula}
A q-pochhammer symbols is defined by
\bel
(a;q)_n=\prod_{i=1}^n (1-aq^{i-1})
\ee
The theta function is defined by
\bel
\th(x;q)=\prod_{n=0}^{\infty} (1-x q^{n} ) (1-x^{-1} q^{n+1}), \quad x \in \mathbb{C}^*, \, |q| <1.
\ee
It satisfies the following relations
\bel
\th (x;q)=\th \left(\frac{q}{x};q \right)=-x\th \left(\frac{1}{x};q \right)
\label{therel}
\ee
From this,  it follows  for $n >0$
\bel
&&\th(q^nx;q)=(-x)^{-n} q^{-\frac{n(n-1)}{2}}  \th (x;q) \\
\label{diffth1}
&&\th(q^{-n}x;q)=(-x)^n q^{-\frac{n(n+1)}{2}}  \th (x;q)
\label{diffth2}
\ee
The elliptic gamma function is defined by
\bel
\Gamma(  x; s,t)=\prod_{j,k=0}^{\infty} \frac{1-x^{-1} s^{j+1} t^{k+1}}{1-x s^{j} t^{k}}, \quad |s|,|t| <1.
\ee
The elliptic gamma function satisfies the following relations.
\bel
&&\Gamma(s x;s,t)=\th (x ;t) \Gamma(x;s,t) \\
&&\Gamma(t x;s,t)=\th (x  ;s) \Gamma(x;s,t) \\
&&\Gamma(s^{-1} x;s,t)=\th^{-1} (x s^{-1};t) \Gamma(x;s,t) \\
&&\Gamma(t^{-1} x;s,t)=\th^{-1} (x t^{-1};s) \Gamma(x;s,t)
\ee

For $j, k \in \Z_{\ge0}$, the above equations lead to the following identities 
\bel
&&\frac{\Gamma(s^j t^k x; s,t)}{\Gamma(  x; s,t)}=(-x)^{-jk} s^{-k \frac{j(j-1)}{2}} t^{-j \frac{k(k-1)}{2}} \prod_{l=0}^{j-1} \theta(s^l x;t) \prod_{m=0}^{k-1} \theta(t^m x;s),  \\
&&
\frac{\Gamma(s^{-j} t^{-k}x;s,t)}{ \Gamma( x;s,t) }= (-x)^{-jk} s^{k\frac{j(j+1)}{2}} t^{j\frac{k(k+1)}{2}}    \prod_{l=1}^{j} \th^{-1} (s^{-l}  x;t) \prod_{m=1}^{k}\th^{-1} ( t^{-m} x;s).  
\label{ellipticID}
\ee

The double-sin function is defined by
\bel
s_{b}(x)=\prod_{j, k=0}^{\infty} \frac{j b+k b^{-1} + Q/2-i x}{j b+k b^{-1} + Q/2+i x}
\ee
which satisfies the relation $s_b(x) s^{-1}_{b}(x)=1$.

%\newpage

\end{document}